\RequirePackage{fix-cm}
\documentclass[smallextended]{svjour3}       
\smartqed  
\usepackage{graphicx}
\usepackage{indentfirst,color}
\usepackage{amssymb}
\usepackage{color}
\usepackage{amsmath}
\usepackage{ulem}
\usepackage{paralist}
\begin{document}
\baselineskip  6mm

\title{Weight hierarchies of a family of linear codes associated with degenerate quadratic forms
\thanks{This research is supported in part by Anhui
Provincial Natural Science Foundation (No. 1908085MA02) and National Natural Science Foundation of China (No. 11701001).}
}
\subtitle{}


\author{Fei Li, Xiumei Li
}


\institute{Fei Li \at
              \email{cczxlf@163.com}           
           \and
Faculty of School of Statistics and Applied Mathematics,
Anhui University of Finance and Economics, Bengbu, {\rm 233041}, Anhui, P.R.China\\
Xiumei Li \at \email{lxiumei2013@mail.qfnu.edu.cn};
Faculty of School of Mathematical Sciences,
Qufu Normal University, Qufu, {\rm 273165}, Shandong, P.R.China\\}

\date{Received: date / Accepted: date}

\maketitle

\begin{abstract}
We restrict a degenerate quadratic form $f$ over a finite field of odd characteristic
to subspaces. Thus, a quotient space related to $f$ is introduced.
Then we get a non-degenerate quadratic form induced by $f$ over the quotient space.
Some related results on the subspaces and quotient space are obtained.
Based on this, we solve the weight hierarchies of a family of linear codes
related to $f.$

\keywords{Weight hierarchy \and Generalized Hamming weight \and Linear code
\and Quadratic form \and Quotient space}
\subclass{ 94B05 \and  11E04 \and 11T71}
\end{abstract}

\section{Introduction}
\label{intro}

Weight hierarchies of linear codes have been an interesting topic for their important value in theory and applications to cryptography.
In 1991, Wei in the paper \cite{20WJ91} presented his wonderful results about weight hierarchies.
It has been shown that the weight hierarchy of a linear code
completely characterizes the performance of the code on the type II wire-tap channel.
Readers can refer to \cite{19TV95} for a detailed survey on the results up to 1995 about weight hierarchies.
The interest towards the knowledge of the weight hierarchy of a linear code has
been continually increasing. Many authors devoted themselves to weight hierarchies of
particular classes of codes \cite{1BL14,2BM00,3CC97,7DF14,11HP98,13JF17,14JL97,21XL16,22YL15}.
In general, it is hard to settle the weight hierarchy of a linear code.

Let $p$ be an odd prime and $ \mathbb{F}_{p^{m}}$ be the finite field with $p^{m}$ elements.
Denote by $ C \subset \mathbb{F}_{p}^{n}$ an $[n,k,d]$ $p$-ary linear code with minimum Hamming distance $d$ \cite{12HP03}.
 Let
$ [C,r]_{p} $ be the set of all $r$-dimensional subspaces of $C.$
For $ V \in [C,r]_{p}, $ define
$$
 Supp(V)=\{i:\ x_{i} \neq 0\ \textrm{for some}\
x = (x_{1}, x_{2}, \ldots , x_{n})\in V\}.
$$
Then we define the $r$-th ($1\leq r \leq k$) generalized Hamming weight $ d_{r}(C)$ of linear code $C$ by
$$
d_{r}(C)=\min\{|Supp(V)|: \ V\in [C,r]_{p}\}.
$$
In particular, $ d_{1}(C)=d.$ The weight hierarchy of $C$ is defined as the set $ \{d_{i}(C): 1\leq i \leq k\}$ (see \cite{11HK92,12HP03,15KJ78}).

Denote by $\mathbb{F}_{p^{m}}^{*}$ the set of nonzero elements in the finite field $\mathbb{F}_{p^{m}}.$
A generic construction of linear code was proposed by Ding et al.(\cite{5DJ15,6DD14}). It is as follows.
Let $\mathrm{Tr}$ denote the trace function from $\mathbb{F}_{p^{m}}$ to $\mathbb{F}_p$
and $ D= \{d_{1},d_{2},\ldots,d_{n}\}\subset \mathbb{F}_{p^{m}}^{*}.$ Define
a linear code $ C_{D} $ with length $ n $ as follows:
\begin{eqnarray}\label{defcode1}
         C_{D}=\{\left( \mathrm{Tr}(xd_1), \mathrm{Tr}(xd_2),\ldots, \mathrm{Tr}(xd_{n})\right):x\in \mathbb{F}_{p^{m}}\},
\end{eqnarray}
and $ D $ is called the defining set. Many classes of linear codes with a few weights
were obtained by choosing properly defining sets \cite{9DL16,17LY14,23YY17,24ZL16,26TX17}.

In this paper, we discuss the generalized Hamming weights of a class of
linear codes $ C_{D}$, whose defining set is chosen to be
\begin{eqnarray}\label{defcode2}
         D=D^{a}_{f}=\{x\in \mathbb{F}_{p^{m}}: \ f(x)=a\}, \ a \in \mathbb{F}_{p}^{*}.
\end{eqnarray}
Here $f$ is a degenerate quadratic form over $\mathbb{F}_{p^{m}}$ with values in $\mathbb{F}_{p}.$

In the paper, we settle the weight hierarchy of $C_{D^{a}_{f}}, \ a\in \mathbb{F}_{p}^{\ast}.$
In our previous work \cite{19LF17}, the weight hierarchy of $C_{D^{a}_{f}} \ (a \in \mathbb{F}_{p}^{\ast}) $
relating to non-degenerate quadratic forms was solved.
In \cite{27WZ94,27WW97}, Z. Wan and X. Wu calculated the weight hierarchies of the projective codes
from quadrics by the theory of finite geometry.
In the case $a=0,$ the weight hierarchy of $C_{D^{a}_{f}}$ can be deduced from Theorem 18 in \cite{27WW97}.

The weight distributions of $C_{D^{a}_{f}}$ have been settled.
In reference \cite{25DD15}, K. Ding and C. Ding constructed the linear
codes $C_{D^{a}_{f}}$ in the case $a=0$ relating to the special quadratic form $\mathrm{Tr}(X^{2})$
and determined their weight distributions.
In \cite{8DW17,24ZL16,24ZF17}, the authors calculated the weight distributions of $C_{D^{a}_{f}}$
for general quadratic forms.
In these articles, it was shown that the linear codes $C_{D^{a}_{f}}$ have a few weights and can be used to get
association schemes, authentication codes, secret sharing schemes with interesting access structures.

Also, by these results, we know that $C_{D^{a}_{f}}$ is an $m$-dimensional linear code.
So we can employ a general formula for calculating the generalized Hamming weights of linear codes defined in (1).
It is given as follows.

\par \vskip 0.2 cm
{\bf Lemma 1.(Theorem 1, \cite{18LF17})}\
For each $ r \ (1\leq r \leq m), $  if the dimension of $ C_{D} $
is $m,$ then $d_{r}(C_{D})= n-\max\{|D \bigcap H|: H \in [\mathbb{F}_{p^{m}},m-r]_{p}\}. $

\par \vskip 0.2 cm
The rest of this paper is organized as follows: in Sect. 2, we present some basic definitions and results
of quadratic forms restricted to subspaces and of induced quadratic forms over quotient spaces of finite fields; in Sect. 3,
using the results in Sect. 2, we give all the generalized Hamming weights of linear codes
defined in (2).

\par \vskip 0.2 cm
\section{Quadratic Form, Dual Space and Quotient Space}

\par \vskip 0.2 cm
\subsection{Restricting Quadratic Forms to Subspaces}

\par \vskip 0.3 cm
The finite field $\mathbb{F}_{p^{m}}$ can be viewed as an $m$-dimensional vector space over $\mathbb{F}_{p}.$
Fix a basis $\mathbf{\upsilon_{1}},\mathbf{\upsilon_{2}},\ldots,\mathbf{\upsilon_{m}} \in \mathbb{F}_{p^{m}}$
and express each vector $X\in \mathbb{F}_{p^{m}}$ in the unique form
$X=x_{1}\mathbf{\upsilon_{1}}+x_{2}\mathbf{\upsilon_{2}}+x_{m}\mathbf{\upsilon_{m}}, $ with
$x_{1},x_{2},\ldots,x_{m}\in \mathbb{F}_{p}.$  We can write
$X=(x_{1},x_{2},\ldots,x_{m})^{T}, $ where $T$ represents the transpose of a matrix.

Let $ f: \mathbb{F}_{p^{m}} \rightarrow \mathbb{F}_{p}$ be a quadratic form
over $\mathbb{F}_{p^{m}}$ with values in $\mathbb{F}_{p}$ \cite{16LN97}.
Set
$$
 F(X,Y)=\frac{1}{2}[f(X+Y)-f(X)-f(Y)].
$$
  We can write $ f(X)=X^{T}AX, $ where $A$ is the
symmetric matrix $(a_{ij})_{1\leq i,j\leq m}$ and $a_{ii}=f(\mathbf{\upsilon_{i}}),
a_{ij}=F(\mathbf{\upsilon_{i}},\mathbf{\upsilon_{j}}).$

The rank $R_{f}$ of quadratic form $f$
is defined to be the rank of matrix $A.$
We say that $f$
is non-degenerate if $R_{f}=m $ and degenerate, otherwise.
We can find a invertible matrix $M$ such that $M^{T}AM$ is a diagonal matrix
$\Lambda=diag(\lambda_{1},\lambda_{2},\ldots,\lambda_{R_{f}},0,\ldots,0).$
Let $\Delta_{f}=\lambda_{1}\cdot\lambda_{2}\cdots\lambda_{R_{f}}.$
When $R_{f}=0, $ we define $\Delta_{f}=1.$
Let $\eta$ be the quadratic
character of $\mathbb{F}_p,$ i.e., $\eta(a)=a^{\frac{p-1}{2}}$ for $a \in \mathbb{F}_p^{*}.$
In the paper, $\eta(0)$ is defined to be zero. Under the congruent transformation of
$ A \rightarrow M^{T}AM, \ \eta(\Delta_{f})$ is an invariant. We called $\eta(\Delta_{f}),$
denoted by $\epsilon_{f},$ the sign of the quadratic form $f.$

For a subspace $H\subseteq \mathbb{F}_{p^{m}}, $ define
$$
H^{\bot}=\{x\in \mathbb{F}_{p^{m}}: \ F(x, y)=0\ \textrm{for each} \ y \in H\}.
$$
Then $H^{\bot}$ is called the dual space of $H.$ And $R_{f}$ can also be defined as the codimension of $\mathbb{F}_{p^{m}}^{\bot}.$
Namely, $ R_{f}+\dim(\mathbb{F}_{p^{m}}^{\bot})=m.$

Let $H$ be a $d$-dimensional subspace of $\mathbb{F}_{p^{m}}.$ Restricting the quadratic form
$f$ to $H,$ we get a quadratic form over $H$ in $d$ variables. It is denoted by $f|_{H}.$
Similarly, we define the dual space $H^{\bot}_{f|_{H}}$ of $H$ under $f|_{H}$ in itself by
$$
H^{\bot}_{f|_{H}}=\{x\in H: \ f(x+y)=f(x)+f(y)\ \textrm{for each} \ y \in H\}.
$$
Let $R_{H}, \epsilon_{H}$ be the rank and sign
of $f|_{H}$ over $H,$ respectively. Obviously, $H^{\bot}_{f|_{H}}=H\bigcap H^{\bot}$ and $R_{H}=d-\dim(H^{\bot}_{f|_{H}}).$

\par \vskip 0.2 cm
\begin{example}
\ Let $f(X)=x^{2}_{1}-2x_{1}x_{2}+x^{2}_{2}$ with $X=(x_{1}, \ x_{2}).$ It is a degenerate quadratic form  over $\mathbb{F}_{p}^{2}.$
After simple calculation, we have $F(X,Y)=x_{1}y_{1}-x_{1}y_{2}-x_{2}y_{1}+x_{2}y_{2},$ where $Y=(y_{1}, \ y_{2}).$
Let $ H=\{(x_{1}, \ x_{2})\in \mathbb{F}_{p}^{2}: \ x_{1} = x_{2}\}.$
It is not hard to get $\mathbb{F}_{p^{m}}^{\bot}=H, R_{f}=1, f|_{H}=0, H^{\bot}_{f|_{H}}=H,  R_{H}=0$ and $\epsilon_{H}=1.$
\end{example}

For $a\in \mathbb{F}_{p},$ the following lemma tells
us the number of solutions in $H$ of the equation $f(x)=a.$

\par  \vskip 0.2 cm
{\bf Lemma 2.(Proposition 1, \cite{19LF17})}\
Let $f$ be a quadratic form over $\mathbb{F}_{p^{m}}$, \  $a\in \mathbb{F}_{p}$ and
$H$ be a $d$-dimensional ($d>0$) subspace of $\mathbb{F}_{p^{m}},$ then the number of solutions of $f(X)=a$ in $H$ is
$$
|H\bigcap D^{a}_{f}|=\left\{\begin{array}{ll}
p^{d-1}+v(a)\eta((-1)^{\frac{R_{H}}{2}})\epsilon_{H}p^{d-\frac{R_{H}+2}{2}}, & \textrm{if\ } \ R_{H}\equiv0 (\textrm{mod}2), \\
p^{d-1}+\eta((-1)^{\frac{R_{H}-1}{2}}a)\epsilon_{H}p^{d-\frac{R_{H}+1}{2}}, & \textrm{if\ } \ R_{H}\equiv1 (\textrm{mod}2),
\end{array}
\right.
$$
where  $v(a)=p-1$ if $a=0,$ otherwise $v(a)=-1.$

{\bf Remark} In Proposition 1 of \cite{19LF17}, $f$ is set to be non-degenerate. In fact, we know from the proof that this condition is unnecessary.
Namely, $f$ can be any quadratic form.

For later use, we need two results in the case that $f$ is a non-degenerate quadratic form. They are listed as follows.

\par \vskip 0.2 cm
{\bf Lemma 3.(Proposition 2, \cite{19LF17})} \ Let $f$ be a non-degenerate quadratic form over $\mathbb{F}_{p^{m}}.$
For each $ r $ with $ 0<2r < m, $ there exist an $r$-dimensional subspace
$H\subseteq \mathbb{F}_{p^{m}}\ (m>2)$ such that $H\subseteq H^{\bot}.$

\par  \vskip 0.3 cm
For $k$ elements
$\beta_{1},\beta_{2},\ldots, \beta_{k}\in \mathbb{F}_{p^{m}},$
the matrix $M(\beta_{1},\beta_{2},\ldots, \beta_{k})$ of them is defined as the $k\times k$ square matrix $(F(\beta_{i},\beta_{j}))_{1\leq i,j\leq k}.$
The discriminant $\Delta(\beta_{1},\beta_{2},\ldots, \beta_{k})$ of them is defined to be $\det(M(\beta_{1},\beta_{2},\ldots, \beta_{k})).$
We denote by $\langle \beta_{1},\beta_{2},\ldots, \beta_{k}\rangle$ the subspace spanned by $\beta_{1},\beta_{2},\ldots, \beta_{k}.$

\par \vskip 0.2 cm
{\bf Proposition 1.}\
Let $f$ be a non-degenerate quadratic form over $\mathbb{F}_{p^{m}}$ and
$H \subset\mathbb{F}_{p^{m}}$ a subspace with $\dim(H\bigcap H^{\bot})=e.$
Then, $\epsilon_{H}\epsilon_{H^{\bot}}= (-1)^{\frac{e(p-1)}{2}}\epsilon_{f}. $

{\bf Proof. } \ Suppose $\dim(H)=r. $ By hypothesis, we can set
$$
H=\langle \alpha_{1},\alpha_{2},\ldots,\alpha_{r-e}, \beta_{1},\beta_{2},\ldots,\beta_{e}\rangle,
$$
$$
H^{\bot}=\langle \gamma_{1},\gamma_{2},\ldots,\gamma_{m-r-e}, \beta_{1},\beta_{2},\ldots,\beta_{e}\rangle,
$$
$$
\langle \alpha_{1},\alpha_{2},\ldots,\alpha_{r-e}, \gamma_{1},\gamma_{2},\ldots,\gamma_{m-r-e} \rangle^{\bot}
=\langle \eta_{1},\eta_{2},\ldots,\eta_{e}, \beta_{1},\beta_{2},\ldots,\beta_{e}\rangle.
$$
We have $\epsilon_{H}=\eta(\Delta(\alpha_{1},\alpha_{2},\ldots,\alpha_{r-e})), \ \epsilon_{H^{\bot}}=\eta(\Delta(\gamma_{1},\gamma_{2},\ldots,\gamma_{m-r-e})).$
And
\begin{align}
\epsilon_{f}
&=\eta(\Delta(\alpha_{1},\alpha_{2},\ldots,\alpha_{r-e},\gamma_{1},\gamma_{2},\ldots,\gamma_{m-r-e}, \beta_{1},\beta_{2},\ldots,\beta_{e},\eta_{1},\eta_{2},\ldots,\eta_{e})) \nonumber \\
&=\eta(\Delta(\alpha_{1},\ldots,\alpha_{r-e},\gamma_{1},\ldots,\gamma_{m-r-e}))\eta(\Delta(\beta_{1},\ldots,\beta_{e},\eta_{1},\ldots,\eta_{e})) \nonumber \\
&=\eta(\Delta(\alpha_{1},\ldots,\alpha_{r-e}))\eta(\Delta(\gamma_{1},\ldots,\gamma_{m-r-e}))\eta((-1)^{e})\eta(\det(M^{2})) \nonumber \\
&=\epsilon_{H}\epsilon_{H^{\bot}}\eta((-1)^{e})\eta(\det(M^{2}_{e})). \nonumber
\end{align}
Here $M_{e}$ is the square matrix
$(F(\beta_{i},\eta_{j}))_{1\leq i, j\leq e}.$
Then the desired result follows and we complete the proof.

\par \vskip 0.2 cm
\subsection{Induced Quadratic Form over a Quotient Space}

\par \vskip 0.3 cm
From now on, we suppose $f$ is a degenerate quadratic form. Let $\overline{\mathbb{F}}_{p^{m}}$ be the quotient space $\mathbb{F}_{p^{m}}/\mathbb{F}_{p^{m}}^{\bot}.$
For $\overline{\alpha} \in \overline{\mathbb{F}}_{p^{m}},$ define $\overline{f}(\overline{\alpha})=f(\alpha).$
It is well-defined. We obtain a non-degenerate quadratic form $\overline{f},$ induced by $f,$ over $\overline{\mathbb{F}}_{p^{m}}.$
Without confusion, we still use $f$ to denote $\overline{f}.$
Let $R_{\overline{f}}$ and $\epsilon_{\overline{f}}$
denote the rank and sign of $f$ over $\overline{\mathbb{F}}_{p^{m}},$ respectively. It is easy to see
$R_{\overline{f}}=R_{f}$ and $\epsilon_{\overline{f}}=\epsilon_{f}.$

\par \vskip 0.1 cm
Since $f$ is a non-degenerate quadratic form over $\overline{\mathbb{F}}_{p^{m}},$ the results in Lemmas 2, 3 and Proposition 1
can be applied to $\overline{\mathbb{F}}_{p^{m}}.$

For $a\in \mathbb{F}_{p},$ set
$$
\overline{D}^{a}_{f}=\{\overline{x}\in \overline{\mathbb{F}}_{p^{m}}: \ f(\overline{x})=a\}.
$$
Obviously, $|\overline{\mathbb{F}}_{p^{m}} \bigcap \overline{D}^{a}_{f}|=|\overline{D}^{a}_{f}|.$ Applying Lemma 2 to $ \overline{\mathbb{F}}_{p^{m}},$, we have
$$
|\overline{D}^{a}_{f}|=\left\{\begin{array}{ll}
p^{\overline{m}-1}+v(a)\eta((-1)^{\frac{R_{f}}{2}})\epsilon_{f}p^{\overline{m}-\frac{R_{f}+2}{2}}, & \textrm{if\ } \ R_{f}\equiv0 (\textrm{mod}2), \\
p^{\overline{m}-1}+\eta((-1)^{\frac{R_{f}-1}{2}}a)\epsilon_{f}p^{\overline{m}-\frac{R_{f}+1}{2}}, & \textrm{if\ } \ R_{f}\equiv1 (\textrm{mod}2),
\end{array}
\right.
$$
where  $\overline{m}=R_{f}=\dim(\overline{\mathbb{F}}_{p^{m}})=m-\dim(\mathbb{F}_{p^{m}}^{\bot}).$

\par \vskip 0.2 cm
\begin{example}
\ Just like Example 1, let $f(X)=x^{2}_{1}-2x_{1}x_{2}+x^{2}_{2}$ with $X=(x_{1}, \ x_{2}),$
a degenerate quadratic form over $\mathbb{F}_{p}^{2}\cong \mathbb{F}_{p^{2}}.$  Because
$\mathbb{F}_{p^{2}}^{\bot}=\{(x_{1}, \ x_{2})\in \mathbb{F}_{p}^{2}: \ x_{1} = x_{2}\},$ $\overline{\mathbb{F}}_{p^{2}}$
is isomorphic to $\{(x_{1}, \ x_{2})\in \mathbb{F}_{p}^{2}: \ x_{1} = -x_{2}\}.$ So, $\overline{f}=4x^{2}_{1}$
is a non-degenerate quadratic form over $\overline{\mathbb{F}}_{p^{2}}.$
It is not hard to get $R_{\overline{f}}=R_{f}=1, \epsilon_{\overline{f}}=\epsilon_{f}=1, $ and
$$
|\overline{D}^{a}_{f}|=\left\{\begin{array}{ll}
1, & \textrm{if\ } \ a=0 , \\
2, & \textrm{if\ } \ \eta(a)=1 , \\
0, & \textrm{if\ } \ \eta(a)=-1 .
\end{array}
\right.
$$
\end{example}

\par \vskip 0.2 cm
Let $\varphi$ be the canonical map from $\mathbb{F}_{p^{m}}$ to $ \overline{\mathbb{F}}_{p^{m}}$ \cite{16L02}. For a subspace $H \subset \mathbb{F}_{p^{m}}, $
denote by $\overline{H}$ the image of $H$ under $\varphi,$ i.e., $\overline{H}=\varphi(H).$
In the absence of confusion, also we use $\overline{H}$ to represent a subspace of $ \overline{\mathbb{F}}_{p^{m}}. $
Let $R_{\overline{H}}$ and $\epsilon_{\overline{H}}$
denote the rank and sign of $f$ over $\overline{H},$ respectively.

\par \vskip 0.2 cm
{\bf Proposition 2.}\ \
Let $H$ be a subspace of $\mathbb{F}_{p^{m}}$ and $\overline{H}=\varphi(H)\subseteq \overline{\mathbb{F}}_{p^{m}},$
then $R_{\overline{H}}= R_{H}, \ \epsilon_{\overline{H}}=\epsilon_{H}. $

\par \vskip 0.2 cm
{\bf Proof. } \ Suppose $\dim(H)=r, \ \dim(H\bigcap H^{\bot})=t.$ Then we set
$$
H\bigcap H^{\bot}=\langle \beta_{1},\beta_{2},\ldots,\beta_{t}\rangle, \ \ H=\langle \alpha_{1},\alpha_{2},\ldots,\alpha_{r-t}, \beta_{1},\beta_{2},\ldots,\beta_{t}\rangle.
$$
So we have
$
\overline{H}=\langle \overline{\alpha}_{1},\overline{\alpha}_{2},\ldots,\overline{\alpha}_{r-t}, \overline{\beta}_{1},\overline{\beta}_{2},\ldots,\overline{\beta}_{t}\rangle.
$

The matrix $M(\alpha_{1},\alpha_{2},\ldots,\alpha_{r-t}, \beta_{1},\beta_{2},\ldots,\beta_{t})$ is the block matrix
$$
\begin{pmatrix}
  M_{1} & O \\
  O & O
\end{pmatrix},
$$
where $M_{1}=M(\alpha_{1},\alpha_{2},\ldots,\alpha_{r-t}).$ Then $R_{H}=Rank(M_{1}), \ \epsilon_{H}=\eta(\det(M_{1})).$

And the matrix $M(\overline{\alpha}_{1},\overline{\alpha}_{2},\ldots,\overline{\alpha}_{r-t}, \overline{\beta}_{1},\overline{\beta}_{2},\ldots,\overline{\beta}_{t})$ is the block matrix
$$
\begin{pmatrix}
  \overline{M}_{1} & O \\
  O & O
\end{pmatrix},
$$
where $\overline{M}_{1}=M(\overline{\alpha}_{1},\overline{\alpha}_{2},\ldots,\overline{\alpha}_{r-t}).$ Then $R_{\overline{H}}=Rank(\overline{M}_{1}), \ \epsilon_{\overline{H}}=\eta(\det(\overline{M}_{1})).$
In fact, $\overline{M}_{1}=M_{1},$ since $f(\overline{x})=f(x)$ for each $x\in \mathbb{F}_{p^{m}}.$ Hence the desired results follow directly and we complete the proof.

\par \vskip 0.4 cm
Define the dual space $\overline{H}^{\bot}$ of $\overline{H}$ by
$$
\overline{H}^{\bot}=\{\overline{x}\in \overline{\mathbb{F}}_{p^{m}}: \ f(\overline{x}+\overline{y})=f(\overline{x})+f(\overline{y})\ \textrm{for each} \ \overline{y} \in \overline{H}\}.
$$
For the dual spaces, we have an interesting conclusion as below.

\par \vskip 0.2 cm
{\bf Proposition 3.}\ \
Let $H$ be a subspace of $\mathbb{F}_{p^{m}},$ then $\overline{H}^{\bot}= \overline{H^{\bot}}. $

\par \vskip 0.2 cm
{\bf Proof. } \ Let $\overline{x} $ be an element of $ \overline{H^{\bot}} $ with $x\in H^{\bot}.$
We have $f(x+y)=f(x)+f(y)$ for each $y\in H.$ So $f(\overline{x+y})=f(\overline{x})+f(\overline{y}).$
Since $\overline{x+y}=\overline{x}+\overline{y}, \ f(\overline{x}+\overline{y})=f(\overline{x})+f(\overline{y}).$
By definition, $\overline{x} \in \overline{H}^{\bot},$ which means $\overline{H^{\bot}}\subset\overline{H}^{\bot}.$
On the other hand, let $\overline{x} $ be an element of $ \overline{H}^{\bot}.$ For each $\overline{y}\in \overline{H},$
we have $f(\overline{x}+\overline{y})=f(\overline{x})+f(\overline{y}).$ So
$f(\overline{x+y})=f(\overline{x})+f(\overline{y})$ and $f(x+y)=f(x)+f(y).$
Thus $x\in H^{\bot}$ and $\overline{x}\in \overline{H^{\bot}}.$ Therefore $\overline{H^{\bot}}\supset\overline{H}^{\bot}.$
In a word, $\overline{H}^{\bot}= \overline{H^{\bot}}. $ The proof is finished.

\section{Weight Hierarchies of Linear Codes Defined in (2)}

By our method, we have successfully settled the weight hierarchies of $C_{D^{a}_{f}}.$
In this case $a=0,$ the weight hierarchies can be derived from Theorem 18 in \cite{27WW97}.
In this section, we will just present the weight hierarchies of $C_{D^{a}_{f}} $ in the case $ a\in \mathbb{F}_{p}^{*}.$

\par  \vskip 0.2 cm
{\bf Theorem 1.}\
Let $f$ be a degenerate quadratic form
over $\mathbb{F}_{p^{m}}$ with rank $R_{f}=2s$ and $a$ a non-zero element in $\mathbb{F}_{p}^{*}.$
Suppose $m=2s+l, \ l=\dim(\mathbb{F}_{p^{m}}^{\bot}),$ then for the linear codes defined in (2), we have
$$
d_{r}(C_{D^{a}_{f}})=\left\{\begin{array}{ll}
p^{m-1}-p^{m-r-1}-((-1)^{\frac{s(p-1)}{2}}\epsilon_{f}+1)p^{s+l-1}, & \textrm{if\ } \ 1\leq r\leq s, \\
p^{m-1}-2p^{m-r-1}-(-1)^{\frac{s(p-1)}{2}}\epsilon_{f}p^{s+l-1}, & \textrm{if\ } \ s \leq r< m, \\
p^{m-1}-(-1)^{\frac{s(p-1)}{2}}\epsilon_{f}p^{s+l-1}, & \textrm{if\ } \  r= m.
\end{array}
\right.
$$

\par \vskip 0.2 cm
{\bf Proof. }\
We will use Lemma 1 to compute $d_{r}(C_{D^{a}_{f}}).$
To do so, we need to know the value of $\max\{|D^{a}_{f} \bigcap H|: H \in [\mathbb{F}_{p^{m}},m-r]_{p}\}.$

$\mathbf{Case}:$  $s\leq r< m.$ If $H_{m-r}$ is an $(m-r)$-dimensional subspace of $\mathbb{F}_{p^{m}},$ then, by Lemma 2, we have
$$
|H_{m-r} \bigcap D^{a}_{f}|\leq 2p^{m-r-1},
$$
and $ |H_{m-r} \bigcap D^{a}_{f}|$ may reach the upper bound $2p^{m-r-1}$ if $R_{H_{m-r}}=1$ or $0.$
We assert that there exists an $(m-r)$-dimensional subspace $H_{m-r}\subset \mathbb{F}_{p^{m}}$
satisfying $R_{H_{m-r}}=1$ and $\epsilon_{H_{m-r}}$ may take values $-1$ or $1.$
Applying Lemma 3 to $\overline{\mathbb{F}}_{p^{m}},$ there is an $(s-1)$-dimensional subspace $\overline{H}_{s-1}\subset \overline{\mathbb{F}}_{p^{m}}$
with $\overline{H}_{s-1}\subset \overline{H}_{s-1}^{\bot}.$ So $\dim(\overline{H}_{s-1}^{\bot})=s+1, \ R_{\overline{H}_{s-1}^{\bot}}=2.$
Applying Lemma 2 to $\overline{\mathbb{F}}_{p^{m}},$ for each $b\in \mathbb{F}_{p}^{*}, \ |\overline{D}^{b}_{f} \bigcap \overline{H}_{s-1}^{\bot}|>p^{s-1}.$
We choose an element $\alpha \in(\overline{D}^{b}_{f}\bigcap \overline{H}^{\bot}_{s-1})\backslash \overline{H}_{s-1}$ and
let $\overline{H}_{s}=\langle \alpha \rangle \bigoplus \overline{H}_{s-1}.$ Then
$\dim(\overline{H}_{s})=s, R_{\overline{H}_{s}}=1$ and the values of $\epsilon_{\overline{H}_{s}}=\eta(b)$ may take $-1$ or $1.$
Note that the hypothesis $ l=\dim(\mathbb{F}_{p^{m}}^{\bot}).$
Therefore, there exists an $(s+l)$-dimensional subspace $H_{s+l}\subset \mathbb{F}_{p^{m}}$ with $\overline{H}_{s+l}=\varphi(H_{s+l})=\overline{H}_{s}.$
Thus the assertion is true since $1\leq m-r\leq s+l. $
By Lemma 2, for $s\leq r< m, $ we have that
$\max\{|D^{a}_{f} \bigcap H|: H \in [\mathbb{F}_{p^{m}},m-r]_{p}\}=2p^{m-r-1}.$

$\mathbf{Case}:$ $1\leq r< s. $ For an $(m-r)$-dimensional subspace $ H_{m-r}\subset \mathbb{F}_{p^{m}},$ we have
$$
\dim(\overline{H}_{m-r})=\dim(H_{m-r}/(H_{m-r}\bigcap \mathbb{F}_{p^{m}}^{\bot}))\geq m-r-l=2s-r.
$$
So, we have $\dim(\overline{H}_{m-r}\bigcap \overline{H}_{m-r}^{\bot})\leq r,$ since $\dim(\overline{H}_{m-r})+\dim( \overline{H}_{m-r}^{\bot})=2s.$ Noting that
$R_{\overline{H}_{m-r}}=\dim(\overline{H}_{m-r})-\dim(\overline{H}_{m-r}\bigcap \overline{H}_{m-r}^{\bot}).$ By Proposition 2, we have $ \ R_{H_{m-r}}=R_{\overline{H}_{m-r}}\geq 2s-2r.$
By Lemma 2, we have
$$
|H_{m-r} \bigcap D^{a}_{f}|\leq p^{m-r-1}+p^{s+l-1},
$$
and $ |H_{m-r} \bigcap D^{a}_{f}|$ may reach the upper bound $p^{m-r-1}+p^{s+l-1}$ if $R_{H_{m-r}}=2s-2r+1$ or $2s-2r.$
We assert that there is such an $(m-r)$-dimensional subspace $ H_{m-r}\subset \mathbb{F}_{p^{m}}$ with $ |H_{m-r} \bigcap D^{a}_{f}|=p^{m-r-1}+p^{s+l-1}.$
By the construction of $\overline{H}_{s}$ as above, we have an $r$-dimensional subspace $\overline{H}_{r}\subset \overline{\mathbb{F}}_{p^{m}}$ satisfying
$R_{\overline{H}_{r}}=1$ and $\epsilon_{\overline{H}_{r}}$ may take values $-1$ or $1.$ And $\dim(\overline{H}_{r}^{\bot})=2s-r, R_{\overline{H}_{r}^{\bot}}=2s-2r+1.$
By Proposition 1, the values of $\epsilon_{\overline{H}_{r}^{\bot}}$ may take  $-1$ or $1,$ too.
Note that $ l=\dim(\mathbb{F}_{p^{m}}^{\bot})$ and $m-r=2s-r+l.$
Thus we can construct an $(m-r)$-dimensional subspace $ H_{m-r}\subset \mathbb{F}_{p^{m}}$
satisfying $ \overline{H}_{m-r}= \overline{H}_{r}^{\bot}.$ Notice that $ \epsilon_{H_{m-r}}=\epsilon_{\overline{H}_{r}^{\bot}}.$
Therefore, $|D^{a}_{f} \bigcap H_{m-r}|=p^{m-r-1}\pm p^{s+l-1}.$ By Lemma 2, we have that
$\max\{|D^{a}_{f} \bigcap H|: H \in [\mathbb{F}_{p^{m}},m-r]_{p}\}=p^{m-r-1}+p^{s+l-1}.$

By Lemma 2, we have $|D^{a}_{f}|=p^{m-1}-\epsilon_{f}(-1)^{\frac{s(p-1)}{2}}p^{s+l-1}.$
Then the desired results follow directly from Lemma 1. And we complete the proof.

\par \vskip 0.2 cm
\begin{example}
Let $(p,m)=(3,4)$ and $f(x)=\mathrm{Tr}(x^{12})=\mathrm{Tr}(x^{3^{2}+3}).$
Then $s=1,\ l=2,\ \epsilon_{f}=1$ and the weight hierarchy of $C_{D^{1}_{f}}$ is $d_{1}=18, d_{2}=30, d_{3}=34, d_{4}=36.$
\end{example}

\par  \vskip 0.2 cm
{\bf Theorem 2.}\
Let $f$ be a degenerate quadratic form
over $\mathbb{F}_{p^{m}}$ with rank $R_{f}=2s+1$ and $a$ a non-zero element in $\mathbb{F}_{p}^{*}.$
Suppose $m=2s+1+l, \ l=\dim(\mathbb{F}_{p^{m}}^{\bot}).$ If $\eta(a)=(-1)^{\frac{s(p-1)}{2}}\epsilon_{f},$
then for the linear codes defined in (2) we have
$$
d_{r}(C_{D^{a}_{f}})=\left\{\begin{array}{ll}
p^{m-1}-p^{m-r-1}, & \textrm{if\ } \ 1\leq r\leq s, \\
p^{m-1}+p^{s+l}-2p^{m-r-1}, & \textrm{if\ } \ s< r< m, \\
p^{m-1}+p^{s+l}, & \textrm{if\ } \ r= m.
\end{array}
\right.
$$

\par \vskip 0.2 cm
{\bf Proof. } \
$\mathbf{Case}:$ $1\leq r\leq s. $ For an $(m-r)$-dimensional subspace $ H_{m-r}\subset \mathbb{F}_{p^{m}},$ we have
$$
\dim(\overline{H}_{m-r})=\dim(H_{m-r}/(H_{m-r}\bigcap \mathbb{F}_{p^{m}}^{\bot}))\geq m-r-l=2s+1-r.
$$
We have $\dim(\overline{H}_{m-r}\bigcap \overline{H}_{m-r}^{\bot})\leq r,$ since $\dim(\overline{H}_{m-r})+\dim( \overline{H}_{m-r}^{\bot})=2s+1.$ Noting that
$R_{\overline{H}_{m-r}}=\dim(\overline{H}_{m-r})-\dim(\overline{H}_{m-r}\bigcap \overline{H}_{m-r}^{\bot}).$ By Proposition 2, we have $ \ R_{H_{m-r}}=R_{\overline{H}_{m-r}}\geq 2s+1-2r.$
By Lemma 2, we have
$$
|H_{m-r} \bigcap D^{a}_{f}|\leq p^{m-r-1}+p^{s+l},
$$
and $ |H_{m-r} \bigcap D^{a}_{f}|$ may reach the upper bound $p^{m-r-1}+p^{s+l}$ if $R_{H_{m-r}}=2s+1-2r.$
We assert that there is such an $(m-r)$-dimensional subspace $ H_{m-r}\subset \mathbb{F}_{p^{m}}$ with $ |H_{m-r} \bigcap D^{a}_{f}|=p^{m-r-1}+p^{s+l},$
which is constructed as follows. Applying Lemma 3 to $ \overline{\mathbb{F}}_{p^{m}},$
we know there is an $r$-dimensional subspace $ \overline{H}_{r}\subset \overline{\mathbb{F}}_{p^{m}}$
with $\overline{H}_{r}\subset \overline{H}_{r}^{\bot}.$ So $\dim(\overline{H}_{r}^{\bot})=2s+1-r,\ R_{\overline{H}_{r}^{\bot}}=2s-2r+1.$
Note that $ l=\dim(\mathbb{F}_{p^{m}}^{\bot})$ and $m-r=2s-r+1+l.$
Thus we have an $(m-r)$-dimensional subspace $ H_{m-r}\subset \mathbb{F}_{p^{m}}$
satisfying $ \overline{H}_{m-r}= \overline{H}_{r}^{\bot}.$ By Proposition 1, we have $\epsilon_{H_{m-r}}=\eta(-1)^{r}\epsilon_{f},$
since $\epsilon_{\overline{H}_{r}}=1, \ \epsilon_{f}=\epsilon_{\overline{f}}$ and $ \epsilon_{H_{m-r}}=\epsilon_{\overline{H}_{m-r}}=\epsilon_{\overline{H}_{r}^{\bot}}.$
Therefore, by hypothesis and Lemma 2, $|D^{a}_{f} \bigcap H_{m-r}|=p^{m-r-1}+p^{s+l}.$ By Lemma 2, we have that
$\max\{|D^{a}_{f} \bigcap H|: H \in [\mathbb{F}_{p^{m}},m-r]_{p}\}=p^{m-r-1}+p^{s+l}.$

$\mathbf{Case}:$ $s< r< m.$ The proof is similar to that of Theorem 1.

By Lemma 2, we have
$|D^{a}_{f}|=p^{m-1}+p^{s+l}.$
Then the desired conclusions follow from Lemma 1. And the proof is completed.

\par \vskip 0.2 cm
\begin{example}
Let $(p,m)=(3,4)$ and $f(x)=\mathrm{Tr}(x^{2}+x^{3+1}).$
Then $s=1,\ l=1,\ \epsilon_{f}=-1$ and the weight hierarchy of $C_{D^{1}_{f}}$ is $d_{1}=18, d_{2}=30, d_{3}=34, d_{4}=36.$
\end{example}

\par  \vskip 0.2 cm
{\bf Theorem 3.}\
Let $f$ be a degenerate quadratic form
over $\mathbb{F}_{p^{m}}$ with rank $R_{f}=2s+1$ and $a$ a non-zero element in $\mathbb{F}_{p}^{*}.$
Suppose $m=2s+1+l, \ l=\dim(\mathbb{F}_{p^{m}}^{\bot}).$ If $\eta(a)=-(-1)^{\frac{s(p-1)}{2}}\epsilon_{f},$
then for the linear codes defined in (2) we have
$$
d_{r}(C_{D^{a}_{f}})=\left\{\begin{array}{ll}
p^{m-1}-p^{m-r-1}-p^{s+l}-p^{s+l-1}, & \textrm{if\ } \ 1\leq r\leq s, \\
p^{m-1}-p^{s+l}-2p^{m-r-1}, & \textrm{if\ } \ s< r< m, \\
p^{m-1}-p^{s+l}, & \textrm{if\ } \ r= m.
\end{array}
\right.
$$

\par \vskip 0.2 cm
{\bf Proof. } \
$\mathbf{Case}:$ $1\leq r\leq s. $ for an $(m-r)$-dimensional subspace $ H_{m-r}\subset \mathbb{F}_{p^{m}},$ we have $R_{H_{m-r}}\geq 2s-2r+1.$
By the corresponding proof of Theorem 2, we know that $\epsilon_{H_{m-r}}=\eta(-1)^{r}\epsilon_{f}$ if $R_{H_{m-r}}= 2s-2r+1.$
By hypothesis and Lemma 2, we have $|D^{a}_{f} \bigcap H_{m-r}|=p^{m-r-1}-p^{s+l}.$

Next we will construct an $(m-r)$-dimensional subspace $ H_{m-r}\subset \mathbb{F}_{p^{m}}$ with $R_{H_{m-r}}= 2s-2r+2$ and
discuss the value of $|D^{a}_{f} \bigcap H_{m-r}|.$ Applying Lemma 3 to $ \overline{\mathbb{F}}_{p^{m}},$
there is an $(r-1)$-dimensional subspace $\overline{H}_{r-1}\subset \overline{\mathbb{F}}_{p^{m}}$
with $\overline{H}_{r-1}\subset \overline{H}_{r-1}^{\bot}.$ So $\dim(\overline{H}_{r-1}^{\bot})=2s-r+2, \ R_{\overline{H}_{r-1}^{\bot}}=2s-2r+3.$
Applying Lemma 2 to $ \overline{\mathbb{F}}_{p^{m}},$ we have, for each $b\in \mathbb{F}_{p}^{*}, |\overline{D}^{b}_{f} \bigcap \overline{H}_{r-1}^{\bot}|>1.$
We choose an element $\alpha \in(\overline{D}^{b}_{f}\bigcap \overline{H}^{\bot}_{r-1})$ and
let $\overline{H}_{r}=\langle \alpha \rangle \bigoplus \overline{H}_{r-1}.$ Then
$\dim(\overline{H}_{r})=r, R_{\overline{H}_{r}}=1$ and the values of $\epsilon_{\overline{H}_{r}}=\eta(b)$ may take $-1$ or $1.$
So $\dim(\overline{H}_{r}^{\bot})=2s+1-r, R_{\overline{H}_{r}^{\bot}}=2s+2-2r$ and
$\epsilon_{\overline{H}_{r}^{\bot}}$ may take values $-1$ or $1,$ too.
Therefore, there exists an $(m-r)$-dimensional subspace $H_{m-r}\subset \mathbb{F}_{p^{m}}$ with $\overline{H}_{m-r}=\overline{H}_{r}^{\bot}.$
By Lemma 2, we have that $|D^{a}_{f} \bigcap H|=p^{m-r-1}\pm p^{s+l-1}.$ Therefore, also by Lemma 2, we have
$\max\{|D^{a}_{f} \bigcap H|: H \in [\mathbb{F}_{p^{m}},m-r]_{p}\}=p^{m-r-1}+ p^{s+l-1}.$

$\mathbf{Case}:$ $s< r< m.$ The proof is similar to that of Theorem 1. We omit the details.

By Lemma 2, we have $|D^{a}_{f}|=p^{m-1}-p^{s+l}.$
Then the desired conclusions follow from Lemma 1. And the proof is completed.

\par \vskip 0.2 cm
\begin{example}
Let $(p,m)=(3,4)$ and $f(x)=\mathrm{Tr}(x^{2}-x^{3+1}).$
Then $s=1,\ l=1,\ \epsilon_{f}=1$ and the weight hierarchy of $C_{D^{1}_{f}}$ is $d_{1}=6, d_{2}=12, d_{3}=16, d_{4}=18.$
\end{example}

Examples 3-5 have been verified by Magma.

\par  \vskip 0.5 cm



\end{document}